\documentclass[twocolumn]{aastex63}
\usepackage{rotating} 
\usepackage{amsmath,amssymb,bbold,mathrsfs}
\usepackage{subfigure}
\usepackage{bm}

\makeatletter
\let\frontmatter@title@above=\relax
\makeatother

\shorttitle{\ion{Mg}{2} h-k lines formation in a rMHD model}
\shortauthors{Afonso Delgado et al.}

\graphicspath{{./}{figures/}}

\begin{document}

\title{Formation of the \ion{Mg}{2} \lowercase{h} and \lowercase{k} polarization profiles in a solar plage model and their suitability to infer magnetic fields.}

\email{dafonso@iac.es}

\author{David Afonso Delgado}
\affil{Instituto de Astrofísica de Canarias, E-38200 La Laguna, Tenerife, Spain}
\affil{Universidad de La Laguna, Dept. Astrofísica, E-38206, La Laguna, Tenerife, Spain}
\author{Tanaus\'u del Pino Alem\'an}
\affil{Instituto de Astrofísica de Canarias, E-38200 La Laguna, Tenerife, Spain}
\affil{Universidad de La Laguna, Dept. Astrofísica, E-38206, La Laguna, Tenerife, Spain}
\author{Javier Trujillo Bueno}
\affil{Instituto de Astrofísica de Canarias, E-38200 La Laguna, Tenerife, Spain}
\affil{Universidad de La Laguna, Dept. Astrofísica, E-38206, La Laguna, Tenerife, Spain}
\affil{Consejo Superior de Investigaciones Científicas, Spain}

\begin{abstract}

 The \ion{Mg}{2} h and k lines are among the strongest in the near-ultraviolet solar spectrum and their line core originates in the upper chromosphere, just below the transition region. Consequently, they have become one of the main targets for investigating the magnetism of the upper solar atmosphere. The recent CLASP2 mission obtained unprecedented spectropolarimetric data of these lines in an active region plage, which have already been used to infer the longitudinal component of the magnetic field by applying the weak field approximation. In this paper, we aim at improving our understanding of the diagnostic capabilities of these lines by studying the emergent Stokes profiles resulting from radiative transfer calculations in a radiative magneto-hydrodynamic (rMHD) time-dependent model representative of a solar plage. To this end, we create a synthetic observation with temporal and spatial resolutions similar to those of CLASP2. We find strong asymmetries in the circular polarization synthetic profiles which considerably complicate the application of the weak field approximation. We demonstrate that the selective application of the weak field approximation to fit different spectral regions in the profile allows to retrieve information about the longitudinal component of the magnetic field at different regions of the model atmosphere, even when the circular polarization profiles are not anti-symmetric and are formed in the presence of strong velocity and magnetic field gradients.

\end{abstract}



\section{Introduction} \label{sec:intro}

The determination of the magnetic field in the outer layers of the solar atmosphere is one of the main challenges in solar physics nowadays and a key requirement for their physical understanding (e.g., the review by \citealt{TrujillodelPino2022}).
In the outer solar atmosphere (i.e., upper chromosphere, transition region, and corona) the density of the solar plasma is low enough for the magnetic pressure to dominate over the plasma pressure ($\beta << 1$, \citealt{PriestBook}) and thus the behavior of this plasma is determined by the magnetic field. The empirical knowledge of the magnetic field is thus key to understand some of the energetic processes that take place in the outer solar atmosphere, like solar flares or the heating of the million degrees solar corona.
 
Atomic spectral lines encode information about the physical properties of the plasma from the regions in the atmosphere where they are formed. In particular, their polarization profiles carry information about the strength and orientation of the magnetic field. While most of the atomic lines observed in the visible and infrared solar spectrum originate in the photosphere and lower chromosphere, a number of them that originate in the upper chromosphere and transition region are resonance lines found in the  ultraviolet (UV) solar spectrum (see the review of \citealt{TrujilloBUeno2017}). In some of these spectral lines, different spectral regions originate at different heights in the atmosphere due to changes in the opacity, a particular case is the \ion{Mg}{2} h and k resonant doublet. While the h and k line centers are formed in the upper chromosphere, near the base of the transition region, their wings originate in the lower layers of the chromosphere. Consequently, the \ion{Mg}{2} h and k doublet is one of the best tools to study both atmospheric regions simultaneously.

By studying the polarization profiles of resonance lines in the UV and, in particular, the fingerprints that the magnetic field introduces on the spectral line polarization, we can thus aim at uncovering the magnetic fields of the upper solar chromosphere. 
There are two key effects induced by the magnetic field, namely, the Hanle and the Zeeman effects. The Hanle effect is the magnetic modification of the scattering polarization in the core of spectral lines, in turn consequence of the radiatively-induced
population imbalances and coherence between the magnetic sublevels of the line's atomic levels (the so-called atomic level polarization). The Zeeman effect is the energy splitting of the degenerate magnetic sublevels of an atomic level and, in turn, of the magnetic components of a spectral line, resulting in spectral line polarization (e.g., \citealt{LL04}). Both effects of course coexist, and their joint action results in the magnetic sensitivity of the scattering polarization in the core and the wings of spectral lines via the Hanle and magneto-optical effects, respectively \cite[see,][]{AlsinaBallester2016,DelPinoAleman2016}. 

Significant challenges, both observational and theoretical, are faced when trying to exploit these capabilities of the UV atomic lines to infer the upper chromosphere magnetic field. UV observations are impossible from ground-based facilities due to the large opacity of the Earth's atmosphere at these wavelengths. Consequently, space telescopes are necessary to observe the Sun at those wavelengths. The first opportunity to study the solar \ion{Mg}{2} h and k lines came with the Ultraviolet Spectrometer and Polarimeter \cite[UVSP,][]{Calvert1979, Woodgate1980} aboard the \textit{Solar Maximum Mission} \cite[SMM,][]{Bohlin1980}. This mission provided the first polarimetric data for \ion{Mg}{2} h and k doublet from a quiet region near the solar limb and showed the first signatures of significant scattering polarization in the far wings of these lines \citep{Hence1987,MansoSainz2019}. Since some years ago, the \textit{Interface Region Imaging Spectrograph} \cite[IRIS,][]{dePontieu2014} has provided systematic observations of the intensity in some UV atomic lines (\ion{Mg}{2} h and k lines among them). Recently, the \textit{Chromospheric LAyer SpectroPolarimeter} \cite[CLASP2,][]{Narukage2016} suborbital mission, launched in 2019, achieved unprecedented spectropolarimetric observations of \ion{Mg}{2} h and k lines in a quiet region near the solar limb and in a plage region. These data allowed to determine, for the first time, the longitudinal component of the magnetic field at several heights going from the upper photosphere to the upper chromosphere in a plage region \citep{Ishikawa2021a}. 

The theoretical and numerical modeling of these strong resonance UV lines is challenging because of all the physical processes involved in their formation. Due to the relatively low density in the higher layers of the chromosphere and the subsequent lower rate of collisional processes, the \ion{Mg}{2} h and k lines form out of thermodynamic equilibrium (NLTE). Moreover, the frequency correlation between the absorbed and the re-emitted photons in scattering processes becomes a key ingredient in the line formation, the so-called partial frequency redistribution (PRD) effects \citep{Milkey1974,Uitenbroek1997}. The formation of the intensity of these lines have been also studied in detail in relatively realistic model atmospheres resulting from 3D radiative magneto-hydrodynamic (rMHD) simulations accounting for 3D radiative transfer (RT) \citep{Leenaarts2013,Leenaarts2013a} and in flare models \citep{Kerr2019,Kerr2019a}. Moreover, when modeling the polarization of these lines, it is also necessary to account for J-state interference between the upper levels of the two transitions (\citealt{Belluzzi2012,DelPinoAleman2016,delPinoAleman2020}).

In this work, we study the formation of the \ion{Mg}{2} h and k lines Stokes profiles in a rMHD simulation representative of a solar plage region and study the suitability of the weak field approximation (WFA) to study the magnetism of the solar chromosphere, and support the results recently presented by \cite{Ishikawa2021a}. In Section \ref{sec:problem} we describe the theoretical problem and the atomic and atmospheric models used in this work. We present the results in Section \ref{sec:results}. We describe some representative Stokes profiles resulting from 1.5D radiative transfer calculations, commenting on the sensitivity of the Stokes profiles to the thermal, dynamic, and magnetic properties along the line-of-sight (LOS). Next, we imitate an observation by degrading our results to the spatial and temporal resolution of the CLASP2 observations. We study the resulting profile and compare it with available observations. Using this synthetic observation we demonstrate the applicability of the weak field approximation to the \ion{Mg}{2} h and k circular polarization profiles to determine the longitudinal component of the magnetic field even when strong gradients in velocity and magnetic field are present in the line formation region. Finally, in Section \ref{sec:conclusions} we present our main conclusions.


\section{Formulation of the problem} \label{sec:problem}

\begin{figure*}[htp!]
\centering
\includegraphics[width=.97\textwidth]{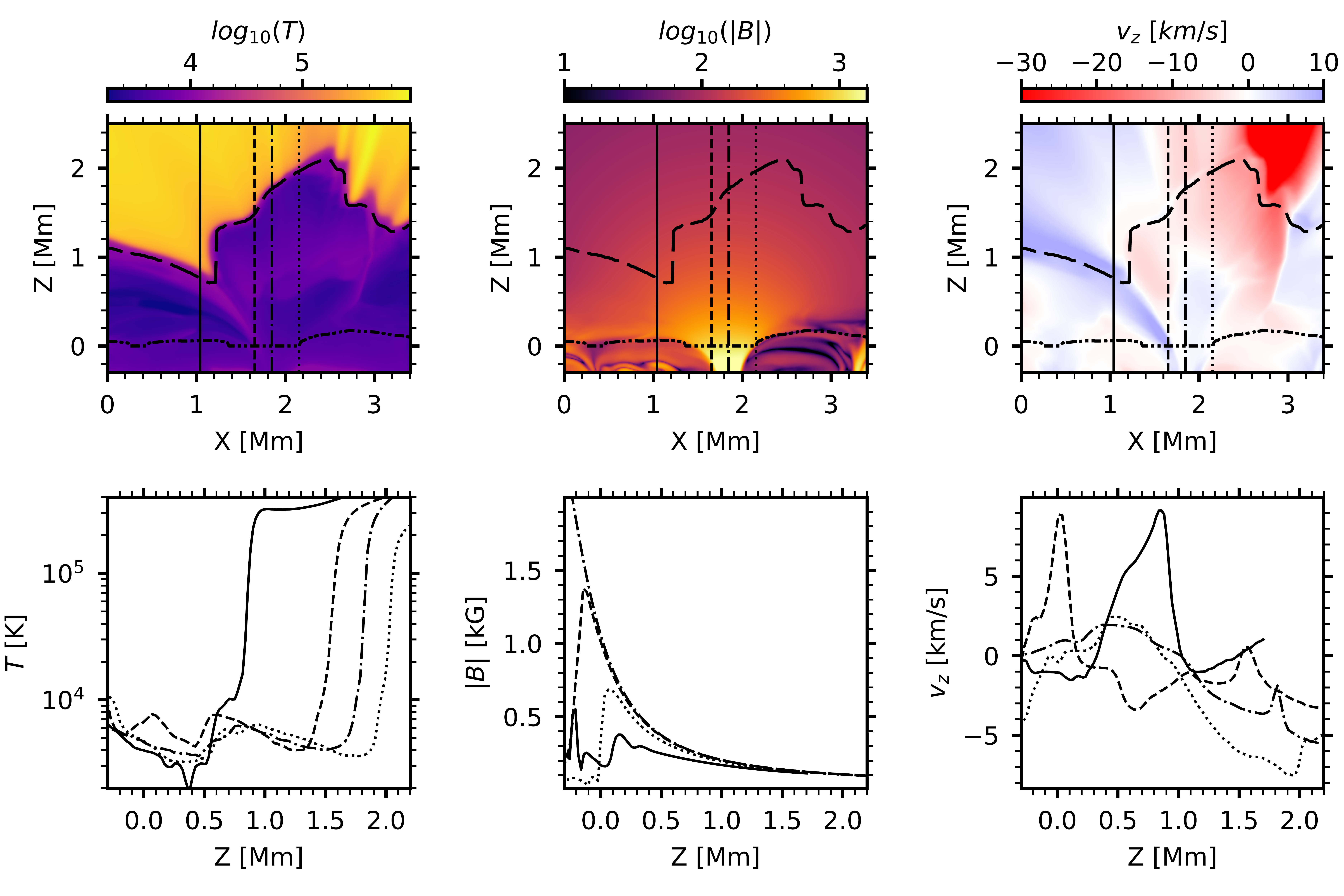}

\caption{Spatial variation (top panels) and vertical stratification (bottom panels) of the temperature (left column), magnetic field strength (middle column), and vertical velocity with $v_z<0$ ($v_z>0$) indicating downflows (upflows) in the atmosphere (right column), in the first snapshot of the rMHD simulation. The different curves in the bottom panels correspond to the locations indicated with the vertical lines in the top panels for the respective line style. The long-dashed curve in the top row panels shows the height at which $\tau=1$ for the \ion{Mg}{2} k line center. The dashed double-dotted curve in the same panels shows the height at which $\tau=1$ in the wings between the \ion{Mg}{2} h and k lines, at 280.00 nm. 
} 

\label{fig:maps}
\end{figure*}

We study the intensity and polarization profiles of the \ion{Mg}{2} h and k doublet and its subordinate triplet in an 2.5D magneto-hydrodynamic atmospheric model representative of a solar plage region, including its temporal evolution for 700~s, with 10~s cadence \citep[see][]{MartinezSykora2020}. With this purpose, we solve the non-local and non-linear 1.5D RT  problem with polarization in a rMHD model. The RT calculations, performed column by column as if they were plane-parallel model atmospheres, take into account NLTE effects, scattering polarization, PRD effects, and J-state interference in the presence of magnetic fields of arbitrary strength (incomplete Paschen-Back regime), and the Doppler shifts produced by the model's macroscopic velocities.
The solution of this RT problem requires the joint solution of the statistical equilibrium equations (SEE)  governing the atomic population and quantum coherence and the RT equations describing absorption and emission processes as the polarized light travels through the solar atmosphere \citep{LL04}. PRD effects are included following the formalism of \cite{Casini2014,Casini2017,Casini2017a}. The RT problem has been solved using the HanleRT code described in \cite{DelPinoAleman2016,delPinoAleman2020}.

The only 3D RT code publicly available today for modeling the polarization of spectral lines without assuming LTE \citep{StepanTrujillo13} is not able to account for all the physical ingredients needed for the modeling of the polarization of the \ion{Mg}{2} h and k lines, namely PRD and J-state interference in the presence of magnetic fields of arbitrary strength. 
While accounting for 3D RT may have an impact on the formation of the Mg {\sc ii} h and k lines, currently it has only been possible to study the effect of horizontal radiation transfer on the intensity profiles \citep{Sukhorukov2017}. These authors concluded that, when comparing horizontal transfer and PRD effects, the latter is of significant more importance for disk-center lines of sight. Moreover, including the horizontal component of the macroscopic velocity allows us to qualitatively account for the lack of axial symmetry inducing linear polarization in 3D model atmospheres \citep{JaumeBestard2021}.
Besides, it is important to note that the considered 2.5D model atmosphere already includes a plane of symmetry which affects the linear polarization and that does not represent the real solar atmosphere. In summary, regarding the main object of study in this paper, namely the fractional circular polarization for the disk center line of sight, we do not expect a significant impact due to the effects of horizontal RT, nor that it would affect our conclusions.

Theoretical studies (e.g., \citealt{Sampoorna2017}) recommend to take into account the angle-dependent frequency redistribution to model the spectral line linear polarization, especially in the presence of a magnetic field. However, the intricate vertical stratification in the columns of the model atmosphere, the need to account for non-axial symmetry due to the presence of non-vertical velocities and magnetic fields, and the amount of calculations that needs to be performed in order to cover an adequate spatial fraction of the model during enough time, results in computing time requirements that are unfeasible at this moment. Consequently, the calculations have been carried out using the angle-averaged approximation \citep{MihalasBook,BelluzziTrujilloBueno2014}.

\subsection{Atomic model} \label{sec:AtomicModel}

In order to study the formation of the \ion{Mg}{2} h and k lines (280.35 and 279.64 nm, respectively), we used an atomic model with three \ion{Mg}{2} terms (five levels) and the ground level of \ion{Mg}{3}. This atomic model is the same used in \citet{delPinoAleman2020}. The \ion{Mg}{2} terms include the ground term, with the single $^2S_{1/2}$ level, and the first and third excited terms with the $^2P_{1/2}$ and $^2P_{3/2}$ levels and the $^2D_{3/2}$ and $^2D_{5/2}$ levels, respectively. The two excited terms in the model are radiatively connected by the s$_{b}$ (279.16 nm) and s$_{r_{a}}$ + s$_{r_{b}}$ (279.88 nm) subordinate transitions. The subordinate triplet can be modeled with the complete frequency redistribution (CRD) approximation \citep{Pereira2015}. 

\subsection{Atmospheric model} \label{sec:AtmosModel}

We model the emergent polarization profiles in a 2.5D rMHD simulation by \cite{MartinezSykora2020}, carried out with the \emph{Bifrost} code \citep{Gudiksen2011}. This model atmosphere includes ion-neutral interaction effects \citep[i.e., ambipolar diffusion;][]{Braginskii1965} and non-equilibrium hydrogen and helium ionization \citep{Leenaarts2011,Golding2014,Golding2016}.

The simulation consists of 70 snapshots covering 700~s (10~s cadence). The computational box is 90~Mm wide, with a uniform horizontal resolution of 4~km, and 43~Mm of vertical extension, from 3~Mm below to 40~Mm above the solar visible surface. The vertical resolution is non-uniform, with a better resolution in the bottom of the model than above (about 12 km between -0.5 and 7~Mm and 70 km at the top of the simulation).

The simulation is representative of an active region plage with opposite magnetic polarities, with a mean magnetic field strength ($|B|$) of $\sim$200~G at the bottom of the photosphere and with a series of magnetic loops connecting both regions. We do not calculate the emergent Stokes profiles for the whole simulation domain, but instead we limit our analysis to a 2~arcsec wide region around the negative polarity, which is dominated by acoustic shocks traveling into the upper layers of the atmosphere. We extract 1D vertical columns from the simulation cutting just above the transition region. We show the plasma properties of this selected region in Fig.~\ref{fig:maps}.



\section{Results: theoretical Stokes profiles} \label{sec:results}

In this section we analyze the Stokes profiles obtained from the RT calculations. First, we show a selection of  profiles calculated at the model resolution to understand how the thermal, dynamic, and magnetic stratification affect the emergent Stokes profiles. Secondly, we create a synthetic observation which imitates the spatial and spectral resolution of the CLASP2 instrument and compare it with some available observations.

\subsection{Stokes profiles at the model resolution} \label{sec:1d}

\begin{figure*}[ht!]
\includegraphics[width=.97\textwidth]{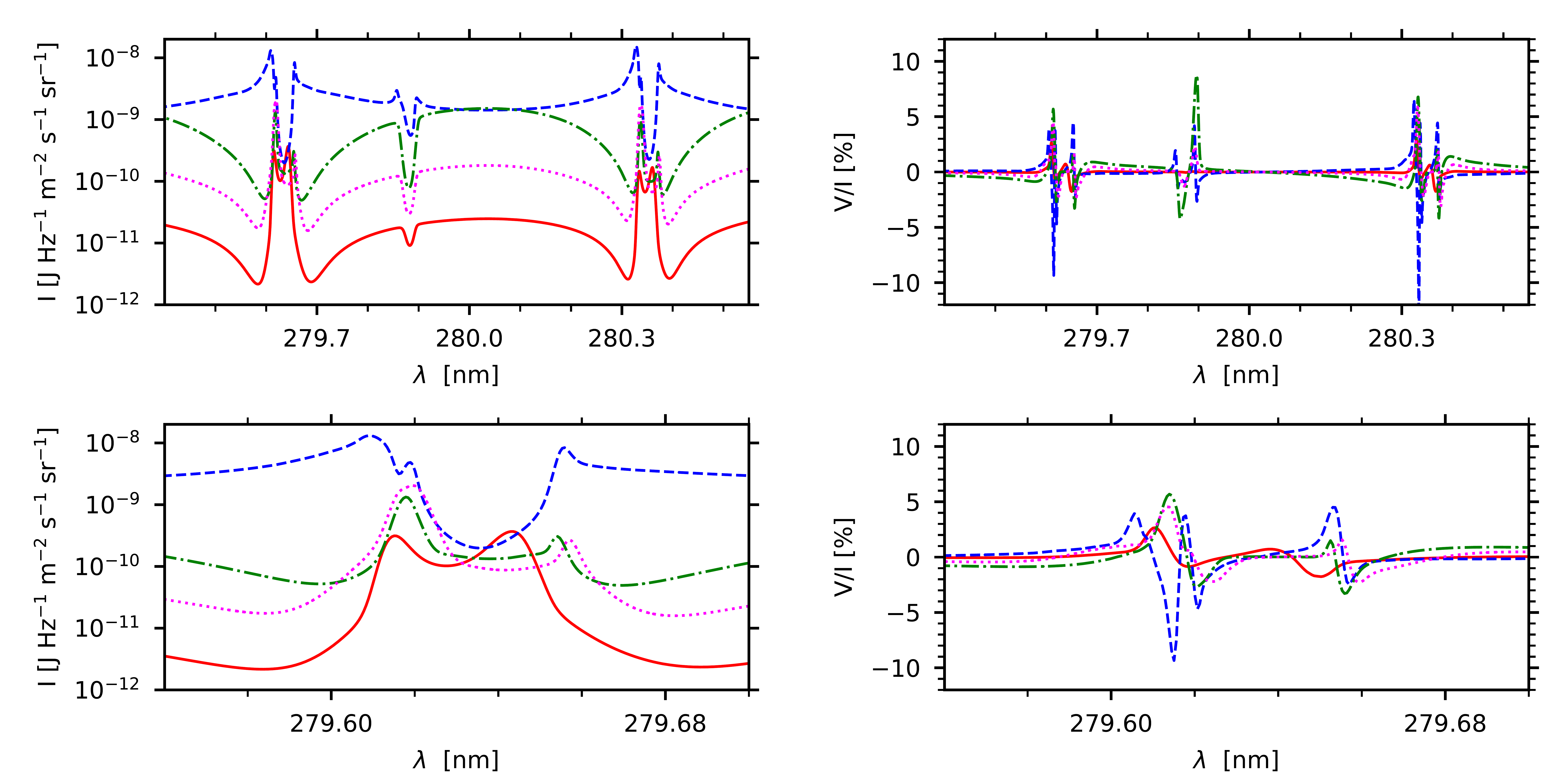}

\caption{Emergent intensity $I$ (left column) and fractional circular polarization $V/I$ (right column) for the four columns indicated in the top row in Figure \ref{fig:maps} (indicated by the style of the line). The bottom row shows only the spectral region around the k line at 279.64 nm.}

\label{fig:profilesComparation}
\end{figure*}

After solving the RT problem in 84 columns in the rMHD model (14 columns in 6 snapshots of the series) we calculate the emergent Stokes profiles for a LOS with $\mu=1$ (with $\mu=\cos{\theta}$ the cosine of the heliocentric angle $\theta$). This selection of columns covers a 2~arcsec region during 60~s in the simulation. In Fig.~\ref{fig:profilesComparation} we show four examples of the intensity and fractional circular polarization profiles emerging from the four selected columns in the model atmosphere (see Fig.~\ref{fig:maps}).

The region of the model we have selected is centered around one of the magnetic concentrations in the model (between 1 and 2.4 Mm in the maps of Fig. \ref{fig:maps}). The plasma properties inside and outside this magnetic concentration show significantly different stratifications (see Fig.~\ref{fig:maps}), leading to Stokes profiles showing different spectral features.

Outside of the magnetic concentration, the intensity profiles show the characteristic double-peaked structure (the peaks are often denoted with k$_{2 v}$ for the violet peak and k$_{2 r}$ for the red peak with respect to the line center for the k line, and h$_{2 v}$ and h$_{2 r}$ equivalently for the h line) with a central depression (often denoted with k$_3$ and h$_3$ for the k and h lines, respectively) in both resonance lines. Moreover, the k line intensity at the k$_{2 v}$ and k$_{2 r}$ peaks is always larger than that in the h line peaks, h$_{2 v}$ and h$_{2 r}$. This behavior is usually found in quiet sun regions both in observations \citep{Schmit2015} and in numerical calculations in semi-empirical atmospheric models \citep{DelPinoAleman2016, delPinoAleman2020}.

The calculated emergent profiles show strong asymmetries due to the LOS velocity gradients. In the first snapshot we always find k$_{2 v} > $k$_{2 r}$ asymmetries (see the magenta dotted curve in Fig. \ref{fig:profilesComparation}, where k$_{2 v}$ is $\approx7$ times larger than k$_{2 r}$). In the corresponding column in the atmospheric model, at the deepest part of the formation region ($\approx 0.5$~Mm above the solar surface) we find an upward velocity of about 2.5~km/s, while at the top of the formation region the velocity is directed downward and is about $-5$~km/s, that is, there is a negative and relatively strong gradient in the region of formation with changing sign of the velocity. The asymmetries with k$_{2 v} > $k$_{2 r}$ seem to appear when there are negative gradients and downward plasma velocities along the line formation region. The opposite case, namely, k$_{2 v} < $k$_{2 r}$, can be found in other snapshots and it corresponds to positive gradients with upward velocity. These results are in agreement with the results of \cite{Leenaarts2013a} in a 3D rMHD model atmosphere. 

\begin{figure*}[ht!]
\includegraphics[width=.97\textwidth]{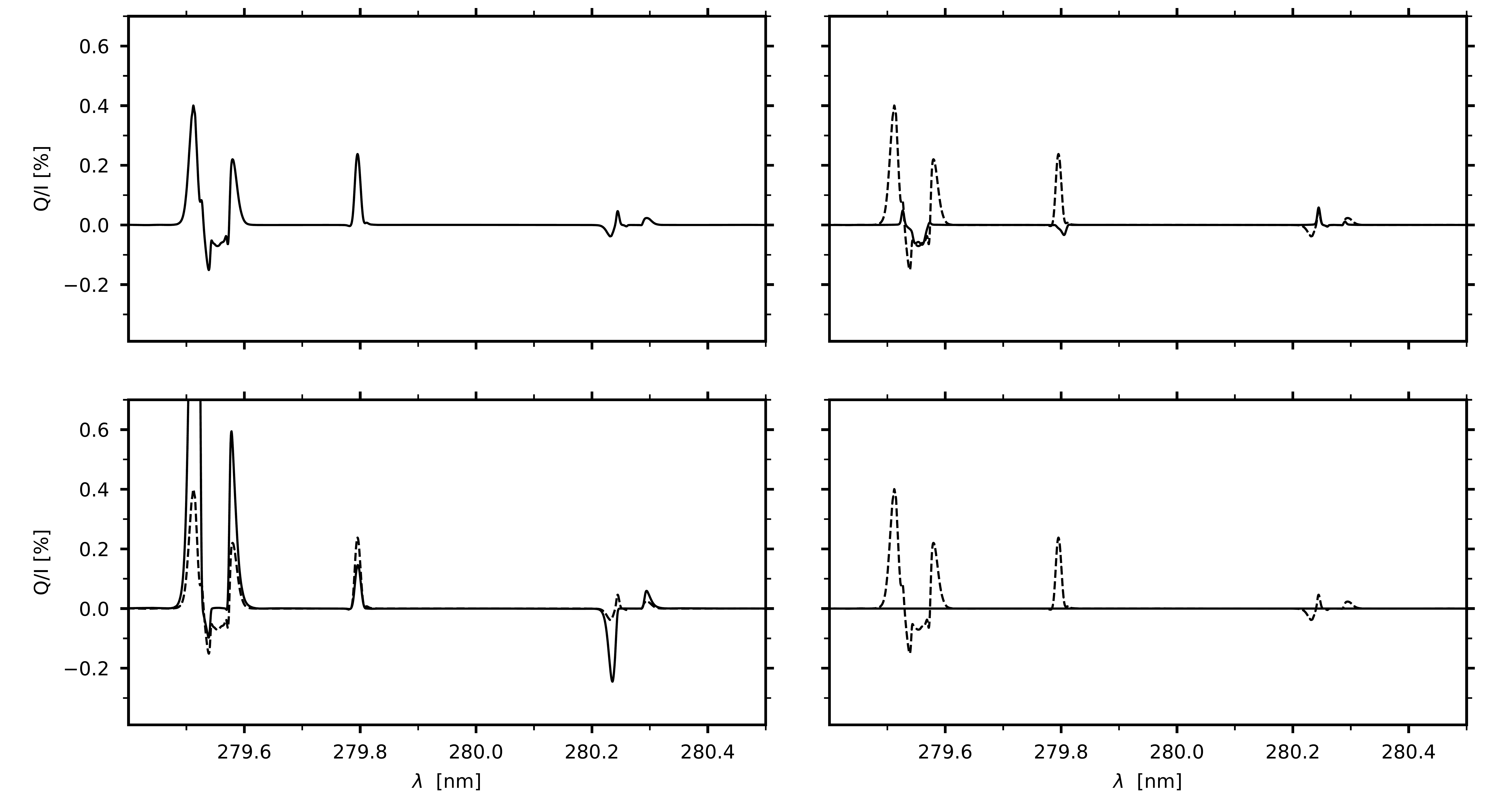}

\caption{Fractional linear polarization $Q/I$ profiles for a specific column in the model atmosphere calculated under four different assumptions: with the magnetic field and the horizontal velocities of the model (top left), with the magnetic field of the model but neglecting horizontal velocities (top right), with the horizontal velocity of the model but neglecting the magnetic field (bottom left), and neglecting both magnetic field and horizontal velocities (bottom right). The dashed curves in the three latter cases show the result from the first case.
}
\label{fig:Qs}
\end{figure*}

In columns close to the center of the magnetic concentration, the intensity profiles show a significant enhancement of the far wings intensity. Consequently, some of these intensity profiles resemble a more typical ``absorption profile'', with k$_3$ and h$_3$ intensities smaller than those in the wings. An example of such a profile can be seen in the blue dashed curve in Fig.\ref{fig:profilesComparation}. The double k$_{2 v}$ peak in this profile is due to the presence of two temperature enhancements at two different heights along the LOS with different associated Doppler shifts (see the dashed curve in Fig.~\ref{fig:maps}). The increased temperature in the lower parts of the atmosphere is responsible for the larger intensity in the far wings.

Regarding the fractional circular polarization $V/I$ (see right column in Fig.~\ref{fig:profilesComparation}), we find the expected four-lobed shape \citep{AlsinaBallester2016,DelPinoAleman2016} in the columns outside of the magnetic concentration, with the outer lobes always showing larger amplitude than the inner lobes, indicative of a negative gradient in the longitudinal component of the magnetic field (i.e., decreasing with height).   
In columns within the magnetic field concentration we find larger polarization signals and more pronounced asymmetries with respect to the line center. Some profiles show more than four lobes (see, e.g., blue profile in Fig.~\ref{fig:profilesComparation}), which correspond to the superposition of different Doppler components (as explained above for the intensity profile corresponding to the same column in the model).

Regarding the fractional linear polarization $Q/I$ and $U/I$ profiles, their amplitudes are comparatively small in the region outside the magnetic field concentration ($<1$\%) and even smaller inside it ($<0.1$\%) for this LOS with $\mu=1$ (disk center observation). To better understand what is the source of this linear polarization\footnote{We remind the reader that for an axially symmetric plane-parallel model atmosphere the only possible source of linear polarization in the emergent radiation for a LOS with $\mu=1$ is the breaking of the axial symmetry due to the horizontal component of the magnetic field or of the macroscopic velocity.} we have performed four numerical experiments: 1) using the magnetic field and the velocity of the model, 2) using the magnetic field of the model, but neglecting the horizontal component of the velocity, 3) using the velocity of the model, but neglecting the magnetic field, and 4) neglecting both the magnetic field and the horizontal component of the velocity. The results are shown in Fig.~\ref{fig:Qs}. As expected, the axially symmetric case (4) results in no fractional linear polarization. Comparing cases 2) and 3), we can see that the horizontal velocity is the main source of scattering linear polarization \citep[for a detailed description of the generation and transfer of scattering polarization in the presence of horizontal velocity gradients, see][]{JaumeBestard2021}. By comparing cases 1) and 3), we can see that the magnetic field mainly depolarizes this scattering linear polarization.

 
\subsection{Stokes profiles at CLASP2 resolution.} \label{sec:avrg}

In order to compare the Stokes profiles of the radiation emerging from this model atmosphere with the observations obtained by the CLASP2 mission we have calculated a synthetic observation integrating the emergent radiation over 2~arcsec and 60~s (similar to the angular resolution and the exposure time of the CLASP2 observations). To emulate a CLASP2 observation we apply both spectral and spatial degradation, convolving with a gaussian of FWHM 0.11~\AA\ and giving more presence to the profiles located at the center of the selected region when we integrate over all the profiles (applying a gaussian of FWHM 1~arcsec), respectively. 

\begin{figure*}[ht!]
\includegraphics[width=.97\textwidth]{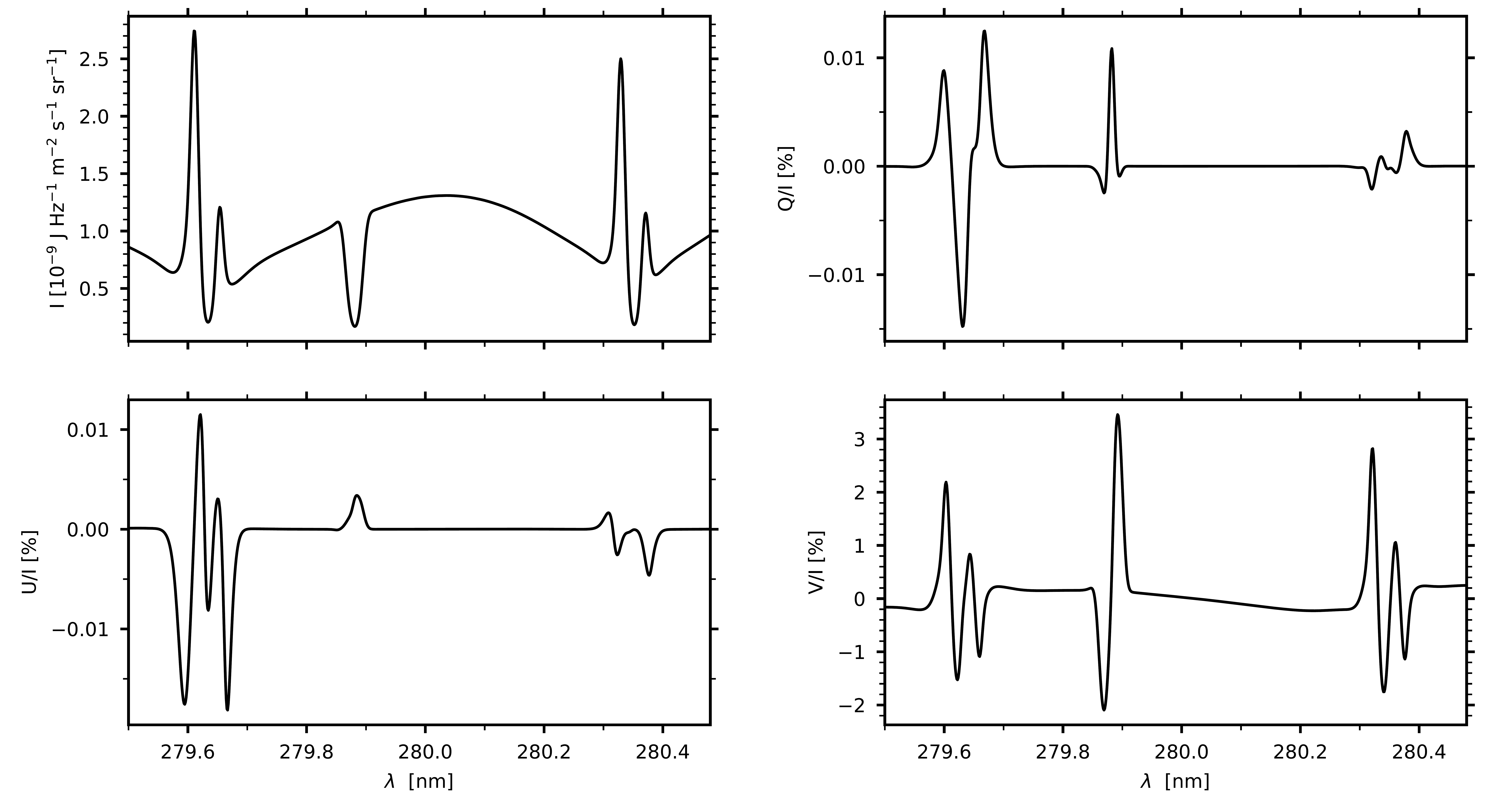}
\caption{Intensity $I$ (top left), fractional linear polarization $Q/I$ (top right) and $U/I$ (bottom left), and fractional circular polarization $V/I$ (bottom right) profiles for a synthetic observation for a LOS with $\mu=1$ emulating the spatial resolution (2~arcsec) and exposure time (60~s) of the CLASP2 observation.
}
\label{fig:meanProfile}
\end{figure*}

The resulting intensity profile shows a significant asymmetry between the two emission peaks, with a relatively deep central depression (k$_{2 v} > $k$_{2 r} > $k$_3$, see Figure \ref{fig:meanProfile}). The dominant profiles are those from inside the magnetic concentration (they have the largest intensity and weight in the spatial convolution) and, therefore, this behavior was to be expected from the analysis in Sect.~\ref{sec:1d}. Likewise, the fractional linear polarization signals are below $\approx0.02$\%, also a consequence of the fact that the dominant profiles are those from inside the magnetic concentration. The fractional circular polarization shows a significant signal instead ($\approx2-3$\% for k and h, respectively). We can see in the fractional circular polarization profile (see bottom right panel of Fig.~\ref{fig:meanProfile}) that the averaged profile still shows the asymmetries between the blue and red lobes that we found inside the resolved profiles in the magnetic concentration (see Sect.~\ref{sec:1d}).

\begin{figure*}[ht!]

\includegraphics[width=.97\textwidth]{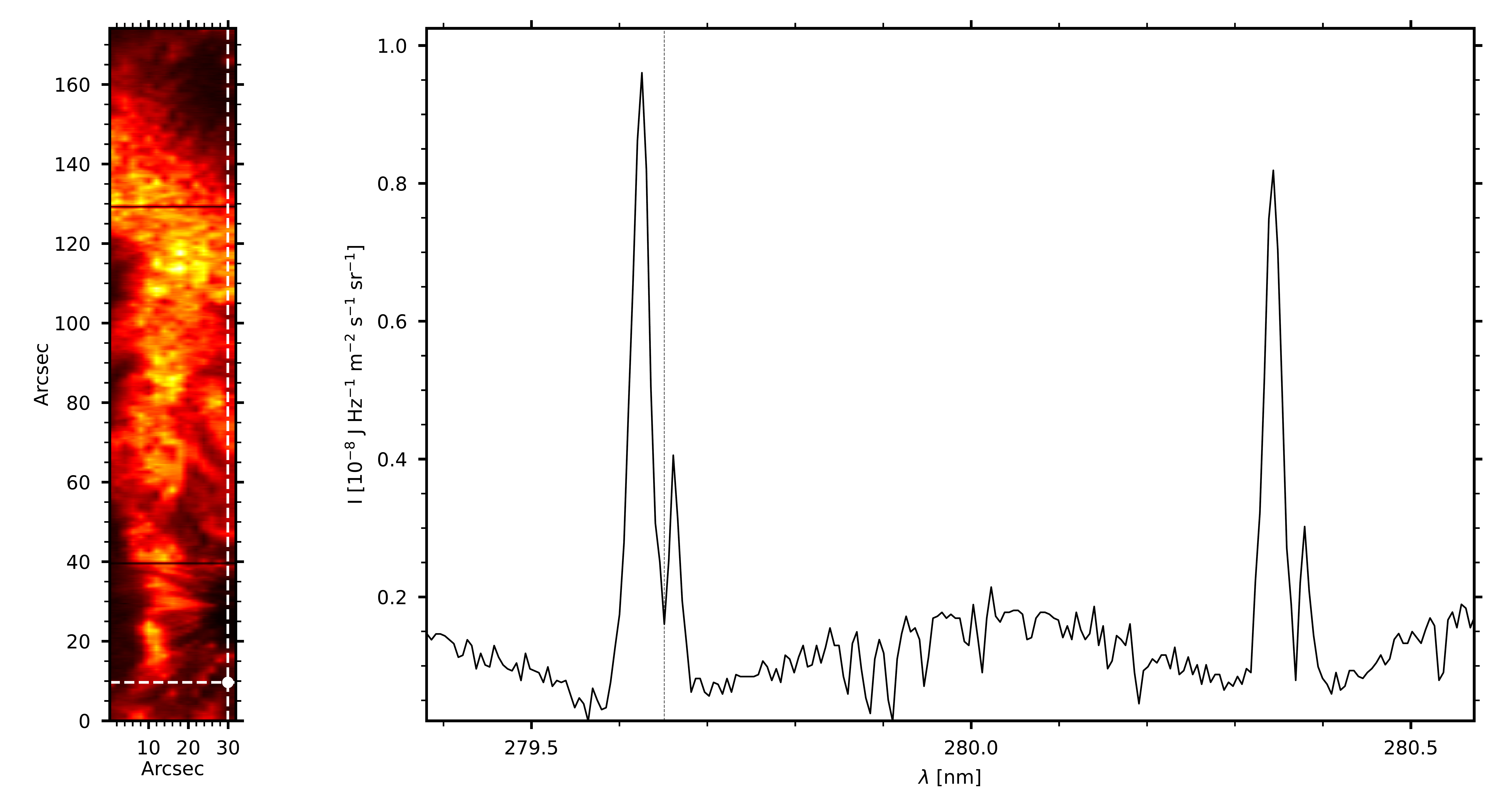}
\caption{Example of an IRIS observation near a magnetic concentration in a plage region. Left panel shows the slit-jaw image at 279.64 nm, and the white dot indicates the position of the profile showed in the right panel. The observation was performed on 2021 October 8 coordinated with the CLASP-2.1 observations. }

\label{fig:iris}
\end{figure*}

Although the theoretical profiles of Fig. \ref{fig:meanProfile} are more complex than the ones we usually find in plage observations \citep[e.g.,][]{Ishikawa2021a}, we have found qualitative similar intensity profiles in the IRIS database. A representative example, with a very deep central depression, asymmetric peaks (with k$_{2v} > $k$_{2r}$), and an enhanced intensity in the far wings, is shown in Figure \ref{fig:iris}. This IRIS observation was carried out in coordination with the CLASP2.1 one. This intensity profile is found close to a plage region and at a magnetic concentration seen as a dark region; we know it is a magnetic concentration from the magnetogram from the same day provided by the Helioseismic and Magnetic Imager on the Solar Dynamics Observatory \citep[SDO/HMI;][]{HMI,HMIpipe}. Likewise, we know that the presence of strong asymmetries in the profiles are indicative of strong gradients in the LOS velocity \citep[][and this work]{Leenaarts2013a}. It thus seems that the profile in Fig. \ref{fig:iris} could correspond to a region in the solar atmosphere with physical conditions akin to those found in the plage simulation.


\section{Results: the Weak Field Approximation suitability} \label{sec:wfa}

Recently \cite{Ishikawa2021a} applied the WFA to the  \ion{Mg}{2} h and k Stokes $I$ and $V$ profiles observed by CLASP2 in order to infer the longitudinal component of the magnetic field ($B_{\rm LOS}$) in the chromosphere of an active region plage. The validity of this approximation when applied to the \ion{Mg}{2} resonant doublet was demonstrated with radiative transfer calculations with constant and stratified magnetic fields in 1D semi-empirical models of the solar atmosphere. However, these semi-empirical models are relatively smooth in their stratification of the physical parameters (temperature, density, and microturbulent velocity) and lack macroscopic velocities. In this section we investigate the validity of the WFA to infer $B_{\rm LOS}$ by applying it to a synthetic observation (Section \ref{sec:avrg}) resulting from radiative transfer calculations in the time-dependent rMHD model described in Section \ref{sec:AtmosModel}, which has a complex stratification of the physical parameters and strong macroscopic velocities with gradients. The $B_{\rm LOS}$ inferred from the WFA can then be directly compared with the magnetic field in the rMHD model.

\subsection{The weak field approximation}

The WFA \citep[][]{LL04} provides an analytical solution to the radiative transfer equations when some fundamental assumptions are satisfied. The first assumption is that the magnetic field strength must be weak enough to guarantee that the Zeeman splitting ($\Delta \lambda_B$) is significantly smaller than the Doppler width of the line ($\Delta \lambda_D$), i.e.:

\begin{equation}
g_{eff}\frac{\Delta\lambda_B}{\Delta\lambda_D} << 1,
\end{equation}

\noindent with $g_{eff}$ the effective Landé factor of the line (1.30 and 1.16 for the case of h and k lines, respectively). $\Delta\lambda_B$ is the Zeeman splitting in \AA:

\begin{equation}
    \Delta\lambda_B = 4.6686\cdot10^{-13}B\lambda_0^2 ,
\end{equation}

\noindent where B is the magnetic field strength in gauss and $\lambda_0$ is the wavelength of the line in \AA, and with $\Delta\lambda_D$ the Doppler width

\begin{equation}
    \Delta\lambda_D = \frac{\lambda_{0}}{c} \sqrt{\frac{2kT}{m}} ,
\end{equation}

\noindent where $c$ is the speed of light, $k$ is the Boltzmann constant, $T$ is the plasma temperature, and $m$ the mass of the atom.

The second assumption is that the longitudinal component of the magnetic field must be constant in the formation region of the spectral line. Then the Stokes $V$ profile is proportional to the longitudinal component of the magnetic field ($B_{\rm LOS}=Bcos\theta_B$, with $\theta_B$ the angle between the field inclination and the LOS) and the first derivative of the intensity profile,
\begin{equation}\label{eq:wfa}
    V = -4.6686\cdot10^{-13}g_{eff}\lambda_0^2B_{\rm LOS}\frac{\partial I}{\partial \lambda}
\end{equation}


\subsection{WFA suitability} \label{sec:wfa}

The spatial extension of the \ion{Mg}{2} h and k formation region makes fitting the whole circular polarization profile with the WFA virtually impossible, as $B_{\rm LOS}$ changes significantly in the region of formation. Consequently, it is necessary to independently fit the inner and outer lobes of these lines in order to obtain a reliable inferred $B_{\rm LOS}$ at the formation heights corresponding to such spectral regions \citep[][]{Ishikawa2021a}. Due to the strong asymmetries present in the circular polarization $V$ profiles calculated in the rMHD model (see bottom right panel of Fig.~\ref{fig:meanProfile}, and compare the red and blue outer lobes), fitting both of them together turns out to also be problematic and we have found that it is impossible to retrieve a reliable $B_{\rm LOS}$ from such a simultaneous fit. For this reason, we decided to apply the WFA independently for the blue and red outer lobes, a decision that can be physically justified as will be shown below. In Fig.~\ref{fig:wfa_fit} we show the application of the WFA to the circular polarization profile obtained in Sect.~\ref{sec:avrg} following this fitting strategy.

\begin{deluxetable*}{ccccc}\label{tab:wfa}
\tablecaption{Inferred $B_{\rm LOS}$ [G] applying the WFA to selected spectral regions of the \ion{Mg}{2} h and k lines. \label{tab:wfa}}
\tablehead{
\colhead{\emph{Line}} & \colhead{Full} & \colhead{Core} & \colhead{Blue wing} & \colhead{Red wing} }
\startdata
$k$ & 312 & 175 & 542 & 373 \\
$h$ & 338 & 177 & 643 & 367
\enddata
\end{deluxetable*}
\vspace{2.5pt}

\begin{figure*}[ht!]
\epsscale{1.0}
\includegraphics[width=.97\textwidth]{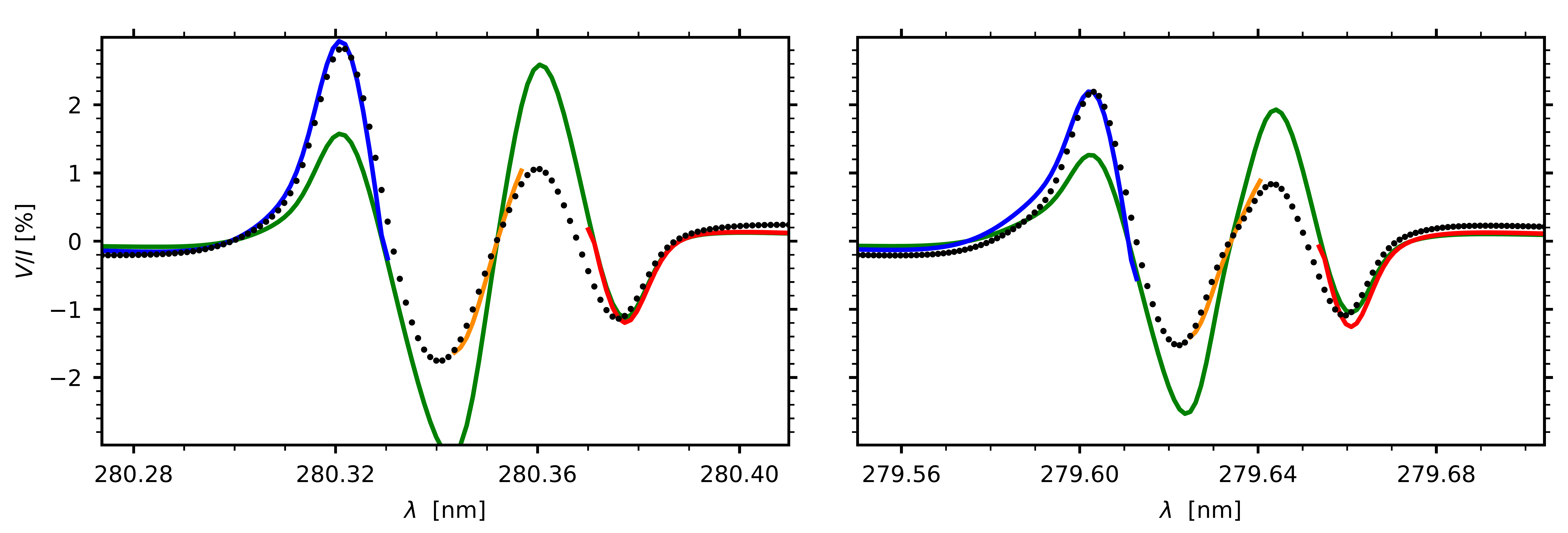}
\caption{Fractional circular polarization $V/I$ profile mimicking a plage region observation (black-dotted curve, see Sect.~\ref{sec:avrg}). The colored solid lines correspond to fits applying the WFA to different spectral regions in the profile, namely the whole spectral range (green), only the inner lobes (orange), the blue outer lobe (blue), and the red outer lobe (red).
}
\label{fig:wfa_fit}
\end{figure*}

\begin{figure*}[ht!]
\centering
\includegraphics[width=0.57\textwidth]{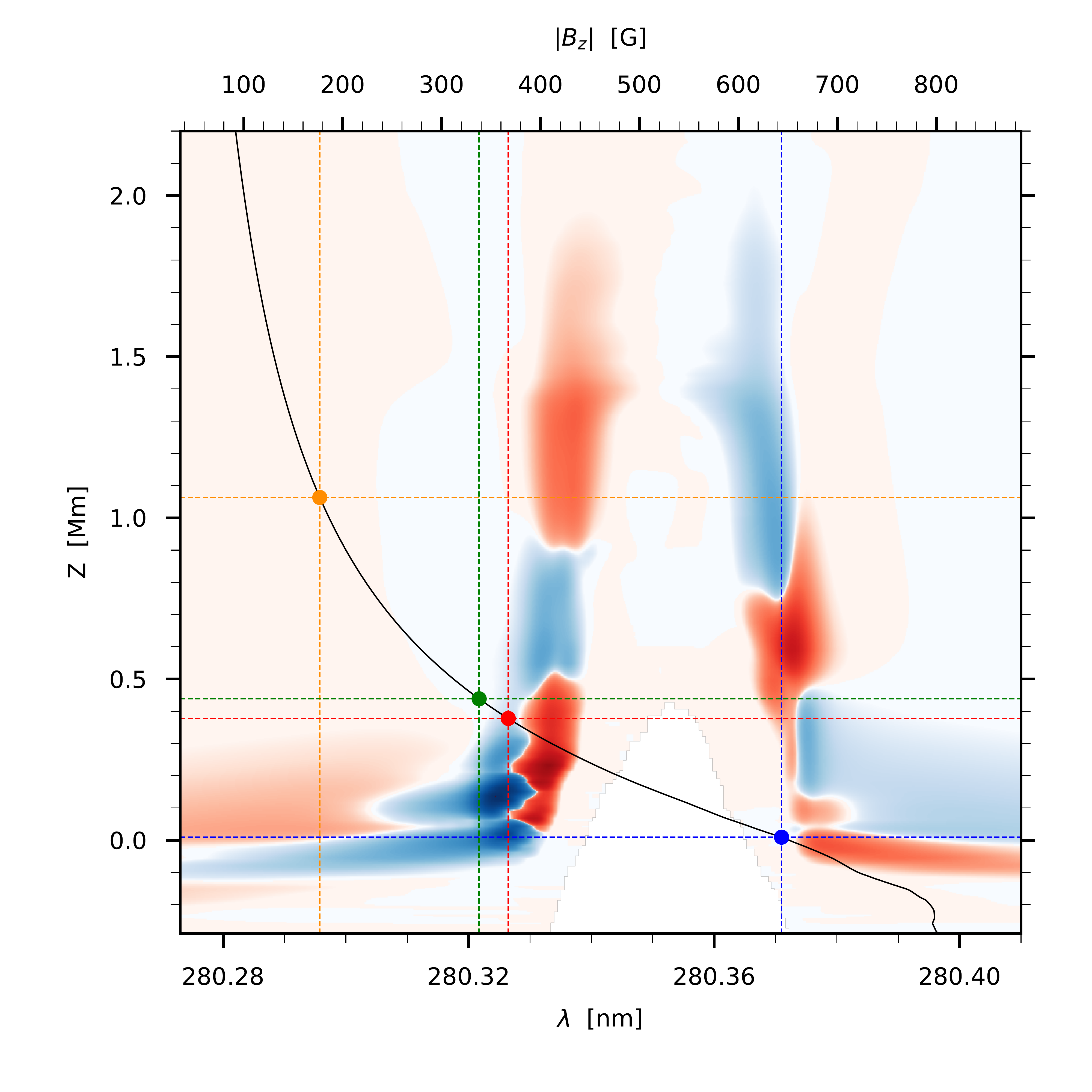}

\caption{Normalized average contribution function to the circular polarization $V$ profile in the rMHD model used in our RT calculations for the \ion{Mg}{2} h line, for each frequency (bottom axis) and height (left axis). The black curve shows the stratification (left axis) of the average longitudinal magnetic field (top axis) over all the columns selected from the rMHD model. The orange, red, blue, and green dots show the value inferred for the longitudinal magnetic field by applying the WFA to the inner lobes, to the red outer lobe, to the blue outer lobe, and to the whole profile, respectively.}

\label{fig:CV_mean}
\end{figure*}

The green curve in Fig.~\ref{fig:wfa_fit} shows the best fit between the circular polarization profile and the first derivative of the intensity when applied to the whole spectral range. It is clear that the circular polarization profile and the first derivative of the intensity are not proportional over the whole range, which is indicative of the stratification of $B_{\rm LOS}$ along the region of formation of the line. However, when distinguishing between inner lobes, blue outer lobe, and red outer lobe, we can find a proportionality constant between the circular polarization profile and the intensity first derivative, thus obtaining three different values for $B_{\rm LOS}$.

In order to understand the meaning of these multiple inferred $B_{\rm LOS}$ values, we compare them with the longitudinal magnetic field found in the atmospheric model. To this end, we have taken the average of $B_{\rm LOS}$ over the same pixels whose emergent Stokes profiles are included in the ``observed'' profile (see Sect.~\ref{sec:avrg}). In addition, we have calculated the contribution function to the circular polarization $V$ profiles for each of those pixels, and took the same average. In Fig.~\ref{fig:CV_mean} we show both averages for the contribution function (for the \ion{Mg}{2} h line, the figure for the k line is almost identical and thus it is not included) and the $B_{\rm LOS}$ stratification, with colored dots marking the $B_{\rm LOS}$ inferred from the WFA from Fig.~\ref{fig:wfa_fit}. The contribution function gives us an idea of how each layer in the model atmosphere (left axis in the figure) contributes to the circular polarization profile at each wavelength (bottom axis). Together with the stratification of $B_{\rm LOS}$ (top axis), we can thus get an idea to the height in the model atmosphere to which the inferred $B_{\rm LOS}$ corresponds.

From the fit to the inner lobes (orange curve and dot in Figs.~\ref{fig:wfa_fit} and \ref{fig:CV_mean}, respectively) corresponding to $B_{\rm LOS}=177$~G, we can find this value at height $\approx1.1$~Mm in the model atmosphere. From the contribution function, we see that the main contribution to the inner lobes indeed mainly comes from $\sim1$-$1.5$~Mm. Regarding the fit to the blue (blue curve and dot) and red (red curve and dot) lobes, corresponding to $B_{\rm LOS}=643$ and $367$~G, respectively, we can find these magnetic fields at heights $\sim0$ and $0.4$~Mm, respectively. From the contribution function we can indeed see that the main contribution to the positive blue lobe (strong blue region between 280.31 and 280.33 nm in the contribution function) is found around $\sim0$-$0.2$~Mm, while the main contribution to the negative red lobe (strong red region between 280.37 and 280.38 nm in the contribution function) is found around $\sim0.5$-$0.8$~Mm. Finally, the fit to the whole profile (green curve and dot) gives us $B_{\rm LOS}=338$~G, which corresponds to a magnetic field in between the maximum and minimum inferred for the different parts of the profile.

This experiment thus demonstrates that the WFA applied to fit different parts of the profile is able to retrieve $B_{\rm LOS}$ values which correspond to the magnetic fields actually found at different heights in the model atmosphere. The velocity gradients found in the model introduce a relative shift of $\sim400$~km between the formation height of the outer lobes, and lead to a relatively complex contribution function (compare with Fig.~S6 in \citealt{Ishikawa2021a}). This complexity is such that we can find opposite contributions to the circular polarization profiles at several wavelength ranges. This multiple contributions are the reason why the fit to the inner lobes in Fig.~\ref{fig:wfa_fit} (orange curve) doesn't really reach the local maxima of those lobes, or why the inferred $B_{\rm LOS}$ for the outer lobes is a bit lower in height than the actual contribution of the correct sign (blue and red colors in the contribution function for the blue and red outer lobes, respectively).

It is clear, however, that we do not have access to this information (contribution function) when working with actual observations. Therefore, without additional information, we cannot really determine to which region of the atmosphere the different $B_{\rm LOS}$ correspond to, besides general descriptions like mid to upper chromosphere (inner lobes) or lower to mid chromosphere (outer lobes). Moreover, the addition of photon noise will of course decrease the accuracy of the inferred $B_{\rm LOS}$.

Nevertheless, the WFA is not only a quick and easy to apply tool to infer $B_{\rm LOS}$ from the \ion{Mg}{2} h and k circular polarization profiles, but the inference it provides is representative of the actual magnetic field which produced such polarization. Moreover, the presence of strong velocity gradients which complicate the formation of the line not only does not impede the application of the WFA, but it can give access to the $B_{\rm LOS}$ in additional layers of the solar atmosphere (although it does not give information about where in the atmosphere).
  
\section{Conclusions} \label{sec:conclusions}
 
In this work we have studied the formation of the polarized spectrum of the \ion{Mg}{2} h and k lines in an rMHD model of a solar plage. To this end, we have performed polarized 1.5D RT calculations in a region around a magnetic concentration in a 2.5D rMHD simulation representative of a solar plage and analyzed the emergent Stokes profiles. These calculations take into account PRD effects as well as J-state interference between the upper levels of the \ion{Mg}{2} h and k transitions. We have combined the emergent Stokes profiles to generate a synthetic observation mimicking the resolution of the CLASP2 observations (2 arcsec of angular resolution and 156 s of exposure time).
 
The selected region of the rMHD simulation shows strong gradients in the vertical (also the LOS) velocity, which results in Stokes profiles with significant asymmetries, confirming previous results in the literature \citep{Leenaarts2013a}. Even though we do not consider horizontal radiation transfer, the horizontal component of the velocity is one of the main contributors to the breaking of axial symmetry (see \citealt{JaumeBestard2021} for a study on the \ion{Ca}{1} line at 4227~\AA). The linear polarization due to this lack of axial symmetry at the disk center LOS is however severely depolarized via the Hanle effect by the strong magnetic fields in this plage model.
 
Most of the intensity profiles emerging from the considered magnetic concentration region, as well as in the synthetic observation, show an asymmetry with k$_{2v} >$ k$_{2r}$ and k$_3$ below the intensity wings between both lines of the doublet (likewise for the h line). Profiles with these properties are not typically found in observations. However, we have found an IRIS observation qualitatively similar to our synthetic observation intensity profile, corresponding to a magnetic concentration in a plage region. We thus expect that the physical conditions of the solar atmosphere corresponding to such observation can be as complex as those found in the rMHD atmospheric model.

For our synthetic observation the fractional linear polarization $Q/I$ and $U/I$ profiles show signals below $\sim0.02$\%. These relatively small signals would be of limited magnetic field diagnostic potential especially in observations affected by photon noise. However, the fractional circular polarization $V/I$ profile shows a remarkable signal above 2\% and thus can provide information on the longitudinal component of the magnetic field $B_{\rm LOS}$.

The main objective of this study has been to test the applicability of the WFA approximation to infer $B_{\rm LOS}$. The WFA approximation had been deemed suitable based on RT investigations in static 1D semi-empirical models of the solar atmosphere with a constant or relatively smooth magnetic field stratification. However, it was not clear how the reliability of the inference could be affected by the strong velocity and magnetic field gradients typical of the solar atmosphere. We could not fit our synthetic observation circular polarization profile with a single $B_{\rm LOS}$, and not even with two different values for the inner and outer lobes, as in previous studies. By studying the contribution function to the circular polarization we have found that the velocity gradients are responsible for this impossibility. In fact, the formation region of the outer lobes of the circular polarization is shifted by about 400~km between them, and thus they map the $B_{\rm LOS}$ at different regions in the model. By fitting individually each of the outer lobes we have been able to infer $B_{\rm LOS}$ at three different regions in the model atmosphere, corresponding to those regions which contribute the most to the circular polarization profile. Therefore, the fit to the inner lobes gave $B_{\rm LOS}$ at the mid-upper chromosphere, the fit to the red outer lobe gave $B_{\rm LOS}$ at the lower-mid chromosphere, and the fit to the blue outer lobe gave $B_{\rm LOS}$ in the photosphere.

Previous studies have shown that the $B_{\rm LOS}$ inferred for the outer lobes of the circular polarization profile by applying the WFA is underestimated by about $\sim13$\% due to PRD effects \citep{Centeno2022,DelPinoAleman2016}. However, this error is considerably smaller than the variation of the magnetic field in the formation region of the outer lobes (see Fig. \ref{fig:CV_mean}) and our conclusions about the suitability of the WFA remain valid.

Consequently, not only the WFA can be applied to the \ion{Mg}{2} h and k circular polarization profiles, even in the presence of strong gradients in the underlying solar atmosphere, but these gradients can give us access to additional information about the $B_{\rm LOS}$ in different layers within the atmosphere. However, the WFA is of course not enough to determine where in the atmosphere the inferred $B_{\rm LOS}$ is located, and alternative methods (e.g., inversions) need to be applied in order to get information about the formation height of the different parts of the Stokes profiles. Moreover, it is clear that photon noise in the circular polarization (not considered in this study because it is not relevant for our conclusions) will reduce the accuracy in the inference with respect to what is shown in this paper.

Due to the significant computational cost of generating a synthetic observation such as the one presented in Sect.~\ref{sec:avrg}, in this work we have focused on a single region (2~arsec) of the rMHD simulation. Nevertheless, we have verified that the conclusions in Sect.~\ref{sec:wfa} are also applicable to the individual 1.5D profiles and, therefore, we can be confident of their validity not being limited to the studied region of this rMHD model. However, it can be interesting to extend this study to other rMHD simulations in the future in order to confirm the generality of our conclusions.

The results of this study confirm the validity of the methodology in \cite{Ishikawa2021a} with regard to the \ion{Mg}{2} lines, which showed the determination of $B_{\rm LOS}$ at several heights in the extended atmosphere of a solar plage. Moreover, they confirm how useful regular spectropolarimetric observations in the spectral region around the \ion{Mg}{2} h and k lines could be for the magnetic diagnostic of the solar chromosphere, and in particular of the circular polarization profile which gives us a quick and relatively simple access to $B_{\rm LOS}$ via the WFA.

\acknowledgements

We are very grateful to Juan Mart\'inez-Sykora (LMSAL) for kindly providing the rMHD time-dependent model used in this investigation,  and for his careful reading of the manuscript. We acknowledge the funding received from the European Research Council (ERC)
under the European Union's Horizon 2020 research and innovation programme (ERC
Advanced Grant agreement No 742265).
 

\bibliography{MgII}
\bibliographystyle{aasjournal}

\end{document}